\documentclass[aps,preprint,floatfix,showpacs]{revtex4}
\usepackage{graphicx}
\begin{document}
\title{Molecular Dynamics in grafted layers of poly(dimethylsiloxane) (PDMS)}
\author{L. Hartmann, F. Kremer\footnote{Corresponding author. e-mail: kremer@physik.uni-leipzig.de, Tel. (-FAX): +49-(0)341-9732-483 (-599)}}
\affiliation{University of Leipzig, Faculty of Physics, Linn\'{e}strasse 5, D-04103 Leipzig, Germany}
\author{P. Pouret, L. L\'{e}ger}
\affiliation{Laboratoire de Physique de la Mati\`{e}re Condens\'{e}e, Coll\`{e}ge de France, 11 Place Marcelin
Berthelot, 75231 Paris Cedex 05, France \vspace{1cm}}
\begin{abstract}\label{abstr}
Dielectric relaxation spectroscopy ($10^{\rm -1}$ Hz to $10^{\rm 6}$ Hz) is employed to study the molecular dynamics of
poly(dimethylsiloxane) (PDMS, $M_{\rm w}=1.7\cdot$10$^{\rm 5}$ g mol$^{\rm -1}$ and $M_{\rm w}=9.6\cdot$10$^{\rm 4}$ g
mol$^{\rm -1}$) as grafted films with thicknesses $d$ below and above the radius of gyration $R_{\rm g}$. For $d<R_{\rm
g}$ the molecular dynamics becomes faster by up to three orders of magnitude with respect to the bulk resulting in a
pronounced decrease of the Vogel temperature $T_{\rm 0}$ and hence the calorimetric glass transition temperature
$T_{\rm g}$. For $d>R_{\rm g}$ the molecular dynamics is comparable to that of the bulk melt. The results are
interpreted in terms of a chain confinement effect and compared with the findings for low molecular weight glass
forming liquids contained in nanoporous glasses and zeolites. Crystallization effects - well known for PDMS - are
observed for films of thicknesses above and below $R_{\rm g}$.
\end{abstract}
\maketitle
\section{Introduction}\label{intro}
\noindent The glass transition and related dynamics of both, polymers
\cite{DutcherACS99,ForrestACIS01,ForrestPSITF00,DKBuch1102} and low molar mass glass-forming liquids
\cite{ArndtPRL97,HuwePRL99,DKBuch602} under the constraint of spatial confinement have been studied extensively by a
variety of experimental techniques. The idea of these studies is to prove the existence of a characteristic,
temperature dependent length scale $\xi$ for the glass transition, i.e. the size of cooperatively rearranging regions
as introduced by Adam and Gibbs \cite{AdamJCP65}. Thus, the dynamics related to the glass transition is expected to
deviate from that of the bulk when the increase of $\xi$ is hindered by the confinement.

\noindent In case of polymers the confinement can be conveniently realized by preparation of thin films via spin
coating, grafting or Langmuir-Blodgett (LB-) techniques. Plenty of experimental investigations has been dedicated to
the determination of the glass transition temperature $T_{\rm g}$ in thin supported or freely standing {\it spin
coated} films by several techniques since the pioneering ellipsometric experiments by Keddie and co-workers
\cite{KeddieEPL94}. A good review of this topic is given in \cite{ForrestPSITF00}. There are only a few experiments
which examine the dynamics related to the glass transition ($\alpha$-relaxation) in thin polymer films in addition to a
mere determination of $T_{\rm g}$. Due to its broad frequency and temperature range, dielectric relaxation spectroscopy
\cite{DKBuch1102,BlumADVM95,FukaoEPL99,FukaoPRE200,FukaoPRE301,FukaoPRE401,HartmannMRC00,HartmannEPJE01} has turned out
to deliver comprehensive information on the changes of the $\alpha$-relaxation but also of secondary relaxations
\cite{FukaoPRE401} when the film thickness is reduced. For the investigation of finite size effects in thin polymer
films it is up to now the {\it only} technique which probes directly molecular fluctuations. Since the capacitance as
measurand can be tailored by variation of the dimensions of the sample capacitor, dielectric spectroscopy does not
suffer sensitivity losses when it is applied to thin polymer films. In some cases dielectric relaxation spectroscopy
can be combined with thermal expansion spectroscopy \cite{FukaoPRE301} to extend the accessible frequency range.
Moreover, second harmonic generation reveals a broadening of the relaxation time distribution in thin supported polymer
films \cite{HallMM97}. Photon correlation spectroscopy indicates a reduction of the $\alpha$-relaxation time in thin
freely standing films of polystyrene \cite{ForrestPRE98}.

\noindent All these studies have revealed deviations of $T_{\rm g}$ in thin films from the bulk value. However, the
sign and the amount of these deviations depend strongly on the interaction with the underlying substrate
\cite{GrohensLMR98,FryerMM00} as it is consistently found for small molecules in untreated pores
\cite{GorbatschowEPL96}. Thus, the conclusion from the observed shifts of $T_{\rm g}$ to the existence of $\xi$ and the
estimation of its value can be done unambiguously only in case of freely standing polymer films for molecular weights
which are sufficiently small to exclude effects of chain confinement \cite{MattsonPRE00}. Moreover, it was emphasized
recently \cite{ReiterEPJE01} that spin coated films have to be considered as a highly metastable form of matter and it
was argued in particular, if alterations of $T_{\rm g}$ can be related to metastable states of spin coated films. This
conjecture suggests to measure the glass transition temperature or to investigate the dynamics in thin polymer films
prepared by other techniques than spin coating such as grafted films or LB-films. Publications concerning this point
are sparse. Optical waveguide spectroscopy yields an identical decrease of $T_{\rm g}$ for thin films of poly(methyl
methacrylate) prepared by spin coating, grafting and by the LB-technique, respectively \cite{PruckerMMCP98}. For
grafted films of hydroxy terminated polystyrene and poly(4-hydroxystyrene) increases of $T_{\rm g}$ of 50 K for a film
thickness of 100 nm are reported \cite{FryerJCP01}.

\noindent In the limit of high molecular weights for freely standing films an additional "sliding motion" of the
polymer chains along their own contour was proposed which leads to an increased chain mobility and hence to a reduction
of $T_{\rm g}$ \cite{deGennesEPJE00}. For thin polymer films the shift of $T_{\rm g}$ is considered as indication of an
underlying distribution of glass transition temperatures, $T_{\rm g}(d)$. This distribution is described by models
which assume the film to be composed of layers whose mobility is enhanced, reduced or equal to that of the bulk.
Dielectric experiments allow to determine the thickness of these layers in thin supported films \cite{FukaoPRE200}.
Moreover, they reveal that the distribution $T_{\rm g}(d)$ in thin films is associated with an increase of the width of
the $\alpha$-relaxation, a reduction of its dielectric strength and a shift of its relaxation rate.

\noindent In this paper, we use dielectric relaxation spectroscopy to investigate the dynamics in thin grafted films of
PDMS and to prove, if effects known from supported spincoated films hold as well for films obtained by different
preparation techniques. The range of film thicknesses covers values above and below the radius of gyration. Evaluation
of the isochronal dielectric loss allows to determine the relaxation rates of the $\alpha$- and the cold
crystallization ($\alpha_{\rm c}$-relaxation), respectively. While the former is related to the glass transition of
amorphous PDMS,  the latter originates from the amorphous fraction after a partial crystallization has occurred. Thus,
from temperature dependence of relaxation rates the Vogel temperature and the thermal expansion coefficient of free
volume are obtained for films of various film thickness and for the bulk material.

\section{Experimental}\label{exp}
\noindent Bulk properties of poly(dimethylsiloxane) (PDMS), such as the glass transition temperatures, the relaxation
processes and their dependence on thermal history have been studied extensively
\cite{ClarsonPolym85,ClarsonPolym96,KirstCPS94,AdachiJPS79}. The preparation of end-grafted layers of PDMS
and the characterization of their structure by small-angle neutron scattering is reported in previous publications
\cite{AuroyMM90,AuroyPRL91,AuroyMM91,LegerAPS99,MarzolinMM01}. In order to form dense end-grafted layers, at first the
aluminium surface of the electrodes has to be chemically modified to prevent the adsorption of monomers while still
keeping the possibility of end-grafting. This modification is accomplished by vapor deposition of a chemically attached
monolayer of short PDMS oligomers (four SiO units) which is terminated by a SiH group. This modified surface is
subsequently incubated with a melt of mono vinyl end terminated PDMS at 120 $^{\rm 0}$C in presence of platinum as a
catalyst. The melt is a mixture of long and short PDMS chains (see Tab. \ref{geom}) where the latter serve to prevent
adsorption of the long chains. Finally, the grafting of the PDMS chains is achieved by a hydrosililation reaction
between the vinyl end groups and the SiH groups of the monolayer. Thus, the long chains are end attached, but not
adsorbed. The grafting density in this procedure is controlled by the concentration of long chains in the mixture
\cite{LegerAPS99}. Chains that have not reacted are rinsed away by toluene, subsequently the layers were dried. Prior
to evaporation of the upper electrodes, the samples have been stored at room temperature for one day in the vacuum
system of the evaporation source. Cast films from a solution of the polymer in toluene with a thickness of some ten
microns were prepared in order to compare the dynamic behaviour of bulk PDMS to that of the grafted polymer. The
solution cast films were dried at ambient temperature in a vacuum system prior to dielectric measurements.

\noindent The thickness ($d_{\rm diel})$ of the grafted samples has been determined from the capacitance of the sample
condenser at the lowest temperature of 120 K and at a frequency of the electric field of 10$^{\rm 5}$ Hz. The value for
the real part $\varepsilon'$ of the dielectric function required for this thickness determination was obtained from
measurements on PDMS bulk samples of a known sample geometry. For some samples the thickness has been measured
independently by ellipsometry ($d_{\rm ellip}$) on dried grafted layers. However, the accuracy of this technique is
limited because of the small difference in the refractive indices of PDMS and the substrate. The knowledge of the
thickness of grafted layers allows to estimate the grafting density $\sigma$ and the distance $b$ between grafting
sites by use of relations derived previously \cite{AuroyMM91}. Values of characteristic quantities of grafted samples
are given in Table \ref{geom}.

\vspace{.5cm}\hspace{5cm}{\it Please insert Table \ref{geom}}\vspace{.5cm}

\noindent Measurements of the dielectric loss $\varepsilon^{\rm ''}$ were performed in the frequency range from
$1\cdot10^{\rm -1}$ Hz to $10^{\rm 7}$ Hz at temperatures between 120 K and ambient temperature. A
Solartron-Schlumberger frequency response analyzer FRA 1260 with a Novocontrol active sample cell BDC-S was used.
Further details on the application of dielectric spectroscopy to study the dynamics in thin polymer films and on the
experimental setup can be found elsewhere \cite{FukaoPRE301,HartmannEPJE01}. In the isothermal representation of
dielectric loss an artifact occurs at high frequencies ($\nu\ge5\cdot10^{\rm 4}$ Hz) due to the finite resistance of
electrodes \cite{BlumADVM95}.

\noindent Since PDMS is known for its tendency to crystallize, the first measurement on each PDMS sample consisted in
recording the dielectric spectra at constant temperatures after the sample has been quenched in liquid nitrogen and
subsequently stored for 30 min at 120 K. The cooling rate of $\approx$50 K min$^{\rm -1}$ obtained during the quenching
procedure is known to be sufficient to avoid crystallization in bulk samples of PDMS \cite{KirstCPS94,AdachiJPS79}.
Therefore, this procedure should allow to monitor the $\alpha$-relaxation in the amorphous polymer. For some samples
dielectric spectra were subsequently recorded during a slow cooling of the sample.

\section{Results and Discussion}\label{resdiss}
\begin{figure}[p]
\begin{center}
\includegraphics[height=18cm,width=12cm]{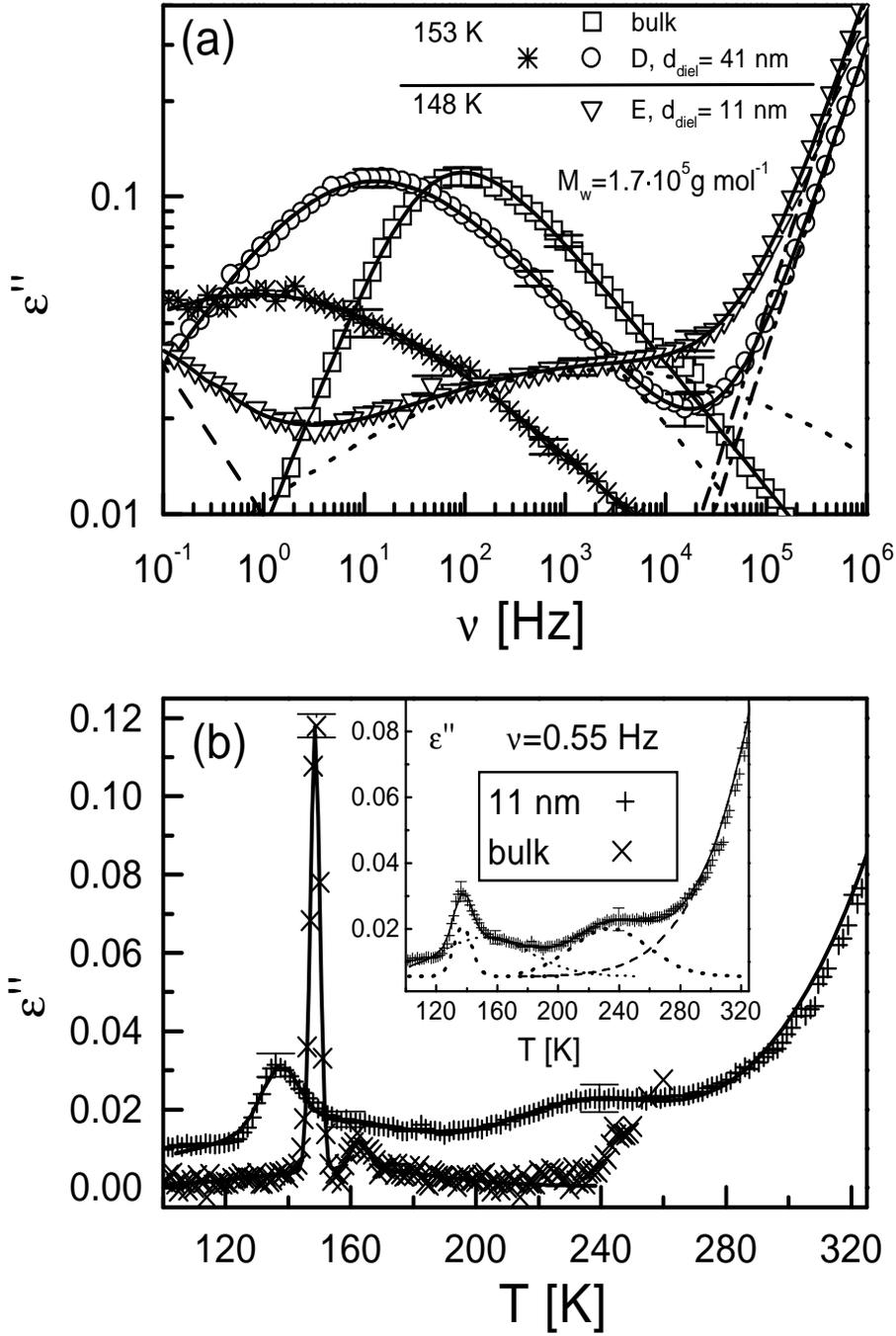}
\caption{(a) Dielectric loss $\varepsilon''$ vs frequency $\nu$ (isothermal representation) for PDMS ($M_{\rm
w}$=170000 g mol$^{\rm -1}$) in the bulk and as grafted layers at $T$=153 K and at $T$=148 K for the the grafted layer
with $d$=11 nm. Solid lines are fits according to the HN-equation (Eq. \ref{havneg}): dotted lines denote relaxation
processes, dashed lines indicate the conductivity and dashed-dotted lines mark an artifact process at high frequencies.
The sample code refers to Table \ref{geom}. Open symbols indicate measurements on samples which were quenched prior to
a heating run. Stars denote measurements on slowly cooled samples. (b) Dielectric loss $\varepsilon''$ vs temperature
$T$ (isochronous representation) for quenched PDMS ($M_{\rm w}$=170000 g mol$^{\rm -1}$) in the bulk ($\times$) and as
grafted layers ($+$) at a frequency $\nu$=0.55 Hz. Solid lines are fits of the data to Eq. (\ref{epsT}). The inset
demonstrates for the grafted layer the decomposition of the fit (dotted lines: Gaussian peaks, dashed line:
conductivity).}\label{fig:1}
\end{center}
\end{figure}

\noindent Figure \ref{fig:1}(a) displays the dielectric loss at $T$=153 K  and $T$= 146 K, respectively as obtained for
the bulk and for thin films of $d$=41 nm ($d>R_{\rm g}$) and $d$=11 nm ($d<R_{\rm g}$) of PDMS ($M_{\rm w}$=170000 g
mol$^{\rm -1}$). Spectra were taken during a heating run of samples which were at first quenched in liquid nitrogen.
Additionally, for $d$=41 nm data are shown which were taken during a slow cooling of the sample. The lines in Fig.
\ref{fig:1}(a) represent fits according to the equation of Havriliak and Negami \cite{HavNeg67} (HN-equation):
\begin{equation}\label{havneg}
\varepsilon^{\rm ''}(\omega;T=const.)=\frac{\sigma_{\rm 0}}{\varepsilon_{\rm 0}}\frac{a}{\omega^{s}}-{\rm
Im}\left[\frac{\Delta\varepsilon}{(1+(i\omega\tau)^{\alpha})^{\gamma}}\right].
\end{equation}
\noindent In this notation, $\Delta\varepsilon$ is the dielectric strength and $\sigma_{\rm 0}$ the DC conductivity.
$\alpha$ and $\gamma$ describe the symmetric and asymmetric broadening of the relaxation peak. The exponent $s$ equals
one for Ohmic behavior, deviations ($s<1$) are caused by electrode polarization. $a$ is a factor having the dimension
$[{\rm s}]^{(1-s)}$. In Fig. \ref{fig:1}(a) at high frequencies an artifact is observed which is treated in the fits as
the low frequency wing of a relaxation process.

\vspace{.5cm}\hspace{5cm}{\it Please insert Figure \ref{fig:1}}\vspace{.5cm}

\begin{figure}[p]
\begin{center}
\vspace{2cm}
\includegraphics{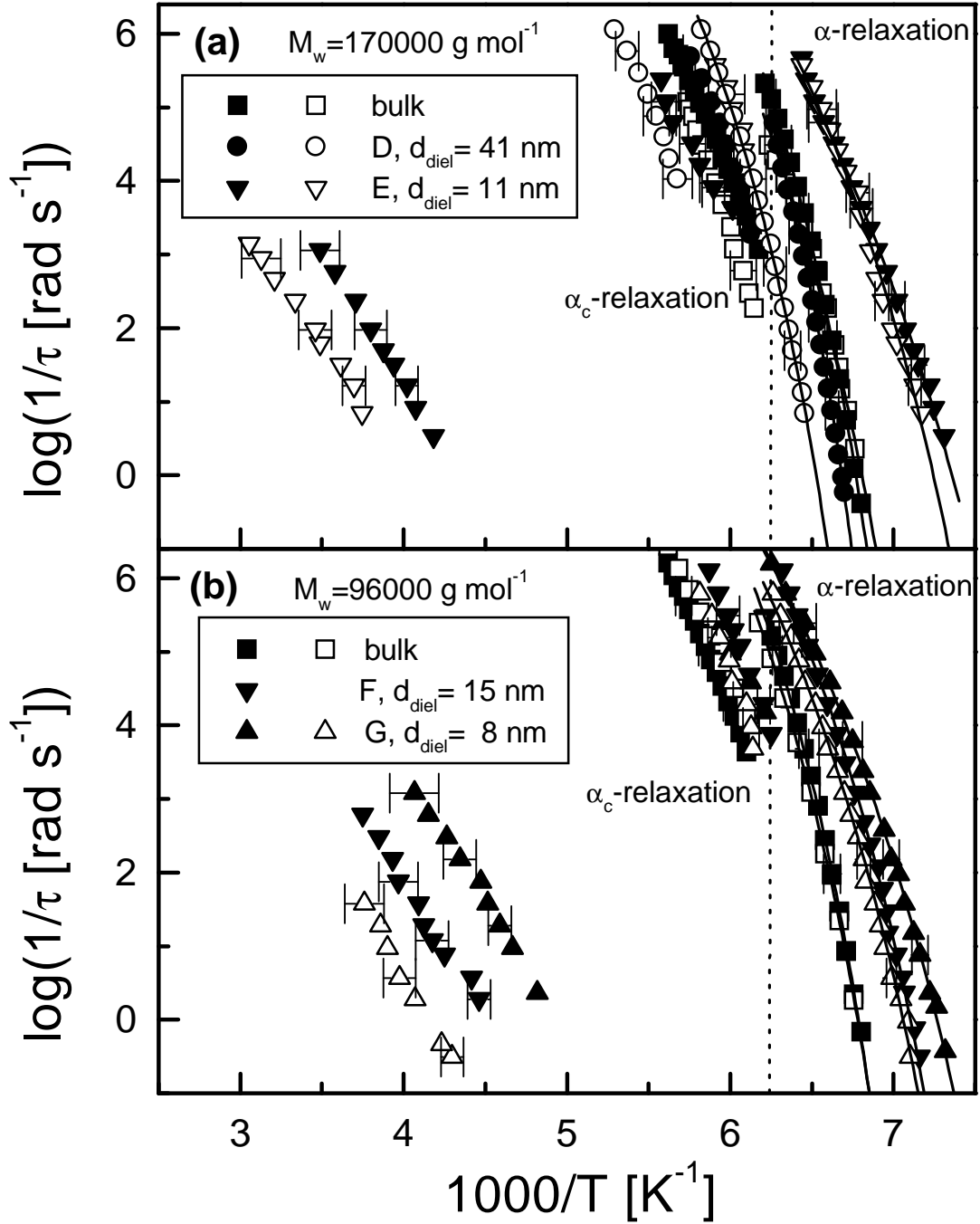}
\caption{Relaxation rate vs inverse temperature (activation plot) for both bulk samples and grafted layers of various
thickness for different thermal treatments of samples and for molecular weights of (a) $M_{\rm w}$=170000 g mol$^{\rm
-1}$ and (b) $M_{\rm w}$=96000 g mol$^{\rm -1}$. The dotted line indicates the bulk crystallization temperature of 162
K. Solid lines are fits according to the VFT-equation (Eq. \ref{vfteq}). Full symbols represent measurements on samples
which were quenched prior to a slow heating, open symbols denote measurements on samples which were cooled down slowly.
The sample code refers to Table \ref{geom}.}\label{fig:2}
\end{center}
\end{figure}

\noindent Figure \ref{fig:1}(a) reveals that in case of quenched samples the $\alpha$-relaxation of the amorphous PDMS
in a film of 41 nm thickness (i.e. a film thickness above the radius of gyration) resembles the $\alpha$-relaxation in
the bulk with respect to the dielectric strength $\Delta\varepsilon$, the shape of loss peak and its width. However,
its maximum position is shifted by about one decade to {\it lower} frequencies with respect to the bulk data. The
comparison of the loss peaks of grafted PDMS ($d$=41 nm) recorded during a heating run after quenching the sample and
during a cooling run shows that in case of the latter the loss peak is again shifted to lower frequencies and that its
dielectric strength has decreased. These findings are due to a partial crystallization of PDMS during the slow cooling
which immobilizes a part of dipoles in crystalline regions and restricts the mobility of those which still contribute
to the $\alpha$-relaxation. The loss peak corresponding to the $\alpha$-relaxation in a grafted PDMS-film with a
thickness of $d$=11 nm (i.e. $d<R_{\rm g}$) clearly differs from that of the thicker film. Fits according to Eq.
(\ref{havneg}) reveal that its maximum is shifted to {\it higher} frequencies by more than 1 decade, the dielectric
strength is reduced and its width is increased with respect to the bulk and the thicker film. However, the analysis of
isothermal data as shown in Fig. \ref{fig:1}(a) suffers, in particular for thin films, from the drawback that the
relaxation processes can be traced only in a limited temperature range due to restrictions given by the conductivity at
low frequencies and the artifact process at high frequencies, respectively. Therefore, the analysis has been extended
by use of the isochronal representation of dielectric loss, i.e. its representation versus temperature at constant
frequencies (Fig. \ref{fig:1}(b)) \cite{FukaoEPL99}. Hence, the dielectric loss can be described as a superposition of
up to three Gaussian functions and a term taking account of the conductivity. The latter is obtained from the
conductivity contribution in Eq. \ref{havneg} under the following assumptions: (i) the temperature dependence of the
DC-conductivity $\sigma_{\rm 0}$ obeys the Vogel-Fulcher-Tammann(VFT)-equation (Eq. \ref{vfteq}) and (ii) the exponent
${\rm s}$ depends linearly on temperature. Thus, the dielectric loss in the isochronal representation can be described
by:
\begin{equation}\label{epsT}
\varepsilon^{\rm ''}(T;\omega=const.)=\sum_{i=1}^{3}\:a_{\rm i}\,\exp\left(\frac{-(T-T_{\rm i}^{\rm max})^{\rm
2}}{w_{\rm i}}\right)+\frac{\sigma_{\rm \infty}}{\varepsilon_{\rm 0}\omega^{\,\rm (m\cdot
T+n)}}\,\exp\left(\frac{-B}{T-T_{\rm 0}}\right)+\lambda,
\end{equation}
where $a_{\rm i}$ and $T_{\rm i}^{\rm max}$ denote the amplitude and the maximum position of the Gaussian functions.
$w_{\rm i}$ corresponds to the width of peak when it has decreased to $1/e$ of its maximum. $\sigma_{\rm \infty}$, $B$
and $T_{\rm 0}$ are the parameters describing the VFT-dependence of the conductivity, $m$ and $n$ are used to describe
the linear dependence of the exponent ${\rm s}$ on the temperature, $\omega=2\,\pi\,\nu$ is the angular frequency of
the electric field and $\lambda$ is an offset. Note that Eq. (\ref{epsT}) is used only to obtain the maximum position
of peaks and hence the activation plot (Fig. \ref{fig:2}, $\tau=1/(2\,\pi\,\nu)$). The results of both fitting
procedures coincide at temperatures for which an analysis according to Eq. \ref{havneg} is possible.

\noindent Examples for fits according to Eq. (\ref{epsT}) are given in Fig. \ref{fig:1}(b) as obtained for the same
data sets of quenched samples as in Fig. \ref{fig:1}(a) at a frequency of 0.55 Hz. The inset exemplifies the
decomposition of the fit into three peaks and the conductivity contribution for the grafted film of $d$=11 nm. For the
quenched bulk sample the isochronal representation of dielectric loss $\varepsilon^{\rm ''}$ is characterized by two
relaxation processes: a symmetric and narrow peak corresponding to the $\alpha$-relaxation of the amorphous PDMS and a
peak corresponding to the $\alpha$-relaxation of the amorphous fraction in the partial crystalline sample ($\alpha_{\rm
c}$-relaxation) which rises above the crystallization temperature of 162 K. The increase at high temperatures for a
frequency of 0.55 Hz is due to the conductivity. This situation holds also for the grafted film with a thickness of 41
nm being larger than $R_{\rm g}$ of the polymer. In contrast, for the film with $d$=11 nm, the two processes are no
longer well separated in temperature, but the $\alpha_{\rm c}$-relaxation appears as shoulder of the
$\alpha$-relaxation. The latter is broadened also in the isochronal representation with respect to the $\alpha$-process
in the bulk. Consistently to Fig. \ref{fig:1}(a), in the isochronal representation the $\alpha$-relaxation in the
grafted film of a thickness of 11 nm is reduced in the maximum of dielectric loss and it is shifted to lower
temperatures with respect to the bulk.

\noindent For the samples E, F and G at high temperatures an additional relaxation process is observed. It is assigned
to segmental fluctuations of low molecular weight PDMS chains ($M_{\rm w}$=5000 g mol$^{\rm -1}$) which are used in the
preparation to decouple the high molecular weight polymers from the surface of the substrate. Compared to any
relaxation process in bulk PDMS this relaxation is by orders of magnitude slowed down presumably due to adhesion to the
surface. The question why such a process is not observed in sample D with $d>R_{\rm g}$ and a grafting density of
$\sigma$=0.14 remains open and requires further studies in dependence on the grafting density. It fits to the
interpretation given above that this segmental fluctuation depends strongly on the thermal history of the sample.

\vspace{.5cm}\hspace{5cm}{\it Please insert Figure \ref{fig:2}}\vspace{.5cm}

\noindent Figure \ref{fig:2} displays the activation plot obtained from fits in the isochronal representation for both
molecular weights under study ((a) $M_{\rm w}$=170000 g mol$^{\rm -1}$, (b) ($M_{\rm w}$=96000 g mol$^{\rm -1}$) and
for different thermal histories of samples as described above. For each molecular weight data for the three relaxation
processes introduced above are shown: the $\alpha$-relaxation, the $\alpha_{\rm c}$-relaxation and a process related to
the dynamics in the interface layer of low molar weight PDMS. Solid lines in Fig. \ref{fig:2} are fits according to the
VFT-equation:
\begin{equation}\label{vfteq}
\log(1/\tau)=\log(1/\tau_{\rm \infty})-\frac{B}{T-T_{\rm 0}},
\end{equation}
where $\tau_{\rm \infty}$ denotes the relaxation time at high temperatures, $B$ is related to the expansion coefficient
of free volume, $\alpha_{\rm f}$, via $\alpha_{\rm f}=B^{\rm -1}$ and $T_{\rm 0}$ is the Vogel temperature.

\noindent The main result in Fig. \ref{fig:2} is a shift of the $\alpha$-relaxation to shorter relaxation rates $\tau$
for films with a thickness distinctly smaller than the radius of gyration. Independent of the sample thickness, the
$\alpha$-relaxation is slightly slowed down in samples which were cooled slowly compared to samples which were quenched
prior to a heating run. However, this effect due to a different thermal history is small in comparison to the shift of
$\tau$ when the film thickness is reduced to values below $R_{\rm g}$ of the polymer. In contrast to these findings,
the $\alpha$-relaxation in the film with $d=41\,{\rm nm}>R_{\rm g}$ is slowed down with respect to the bulk (open and
full circles in Fig. \ref{fig:2}(a)). This effect is even more pronounced when the sample is cooled slowly. The
$\alpha_{\rm c}$-relaxation shows no dependence on the film thickness which could be separated from influence of
thermal history of the samples. Thus, the results indicate that for films in the regime of {\it chain confinement}
(i.e. $d<R_{\rm g}$) the $\alpha$-relaxation is distinctly faster than in the bulk, whereas for films with $d>R_{\rm
g}$ the dynamics resembles well to that in the bulk. The observed slowing down of the $\alpha$-relaxation in thick
films is attributed to interactions with the electrode which prevail the confinement effect at this film thickness. In
films with $d>R_{\rm g}$, the transition between the $\alpha$- and $\alpha_{\rm c}$-relaxation occurs, as in the bulk,
at a well defined temperature ($T\approx 162$ K), whereas no such sharp transition can be found in films with
thicknesses below $R_{\rm g}$. In these films the $\alpha_{\rm c}$-relaxation appears as a shoulder of the
$\alpha$-relaxation whose maximum is poorly defined. Thus, it can be only stated that the transition temperature
between the $\alpha$- and $\alpha_{\rm c}$-relaxation is the same for all films with an uncertainty of $\pm 5K$.

\begin{figure}[p]
\begin{center}
\vspace{2cm}
\includegraphics{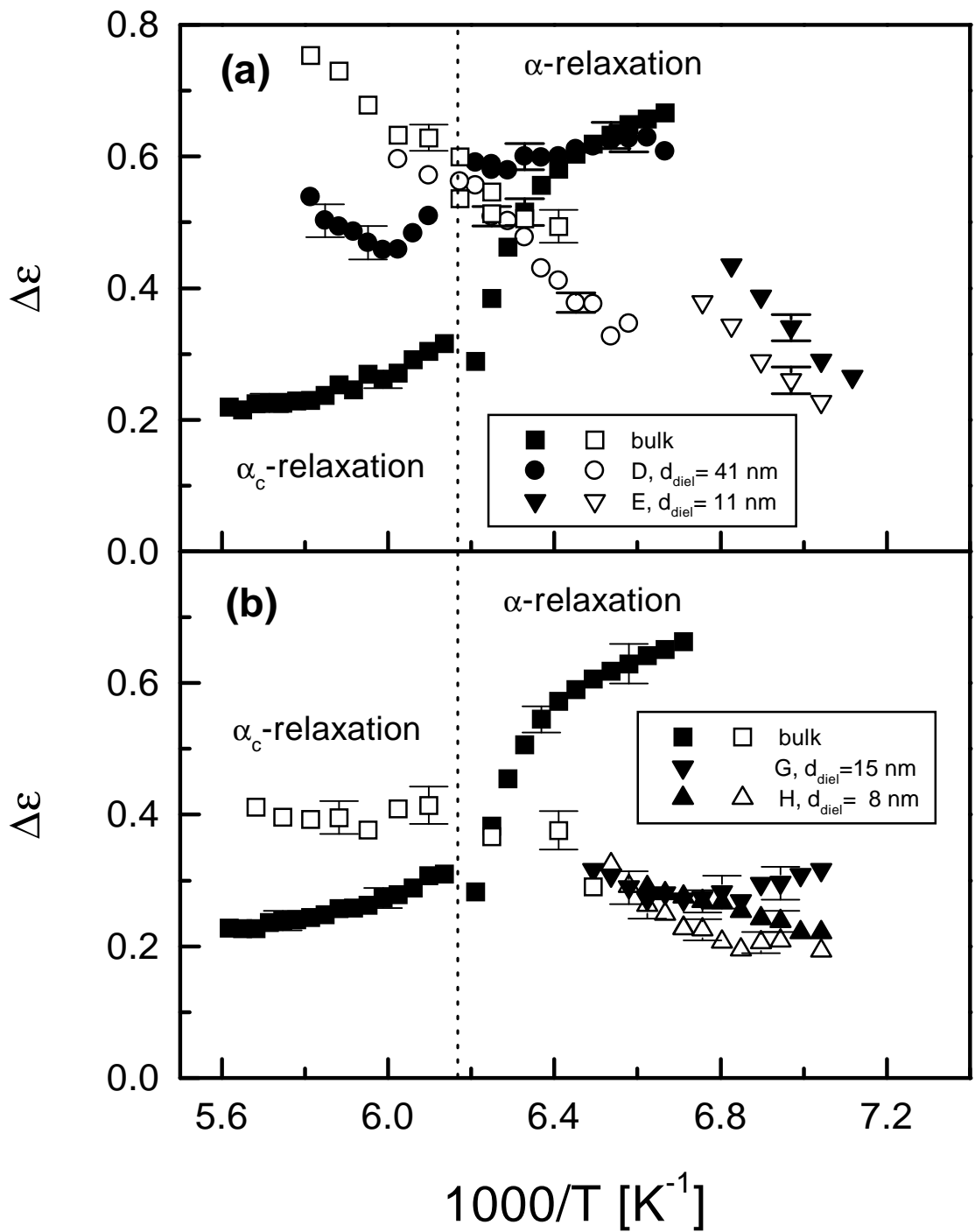}
\caption{Dielectric strength $\Delta\varepsilon$ as obtained from HN-fits vs inverse temperature for molecular weights
of (a) $M_{\rm w}$=170000 g mol$^{\rm -1}$ and (b) $M_{\rm w}$=96000 g mol$^{\rm -1}$. The notation is the same as in
Fig. \ref{fig:2}, the sample code refers to Table \ref{geom}.}\label{fig:3}
\end{center}
\end{figure}

\vspace{.5cm}\hspace{5cm}{\it Please insert Figure \ref{fig:3}}\vspace{.5cm}

\noindent Figure \ref{fig:3} displays the dielectric strength $\Delta\varepsilon$ as obtained from fits of Eq.
\ref{havneg} to isothermal data. For quenched bulk samples, $\Delta\varepsilon$ of the $\alpha$-relaxation is
characterized by a gradual decrease when the temperature approaches the phase transition at 162 K from below. Due to
partial crystallization, $\Delta\varepsilon$ of the $\alpha_{\rm c}$-relaxation is notedly reduced with respect to the
$\alpha$-relaxation. For bulk samples which were cooled slowly, the temperature dependence of $\Delta\varepsilon$ is
given by a steady decrease when the temperature is lowered. The same situation as in the bulk is found for the grafted
film of 41 nm thickness with the only difference that $\Delta\varepsilon$ of the $\alpha_{\rm c}$-relaxation increases
with increasing temperature after the sample has been quenched. Here, the determination of $\Delta\varepsilon$ is
complicated due to the separation of the high frequency artifact especially at the highest temperatures.

\noindent In thin films with a thickness smaller than $R_{\rm g}$ of the polymer a reduction of the dielectric loss
$\varepsilon^{\rm ''}$ has been detected as shown in Fig. \ref{fig:1}(a). This leads to a decrease of
$\Delta\varepsilon$ of the $\alpha$-relaxation by about 50\% to 60\%. The dielectric strength of the $\alpha_{\rm
c}$-relaxation can not be determined from HN-fits according to Eq. (\ref{havneg}) since this process is too weak and
appears only as a shoulder of the $\alpha$-relaxation in isochronal representation of dielectric loss (Fig.
\ref{fig:1}(b)). The decrease of $\Delta\varepsilon$ occurs also in grafted layers after they were quenched ruling out
crystallization as the cause for this reduction. It is instead assumed that the reduction of the dielectric strength of
the $\alpha$-relaxation in films with $d<R_{\rm g}$ is due to a restriction of cooperative fluctuations or a partial
immobilization of dipoles in interfaces near the electrodes \cite{FukaoPRE301,HartmannEPJE01}.

\begin{figure}[p]
\begin{center}
\vspace{2cm}
\includegraphics{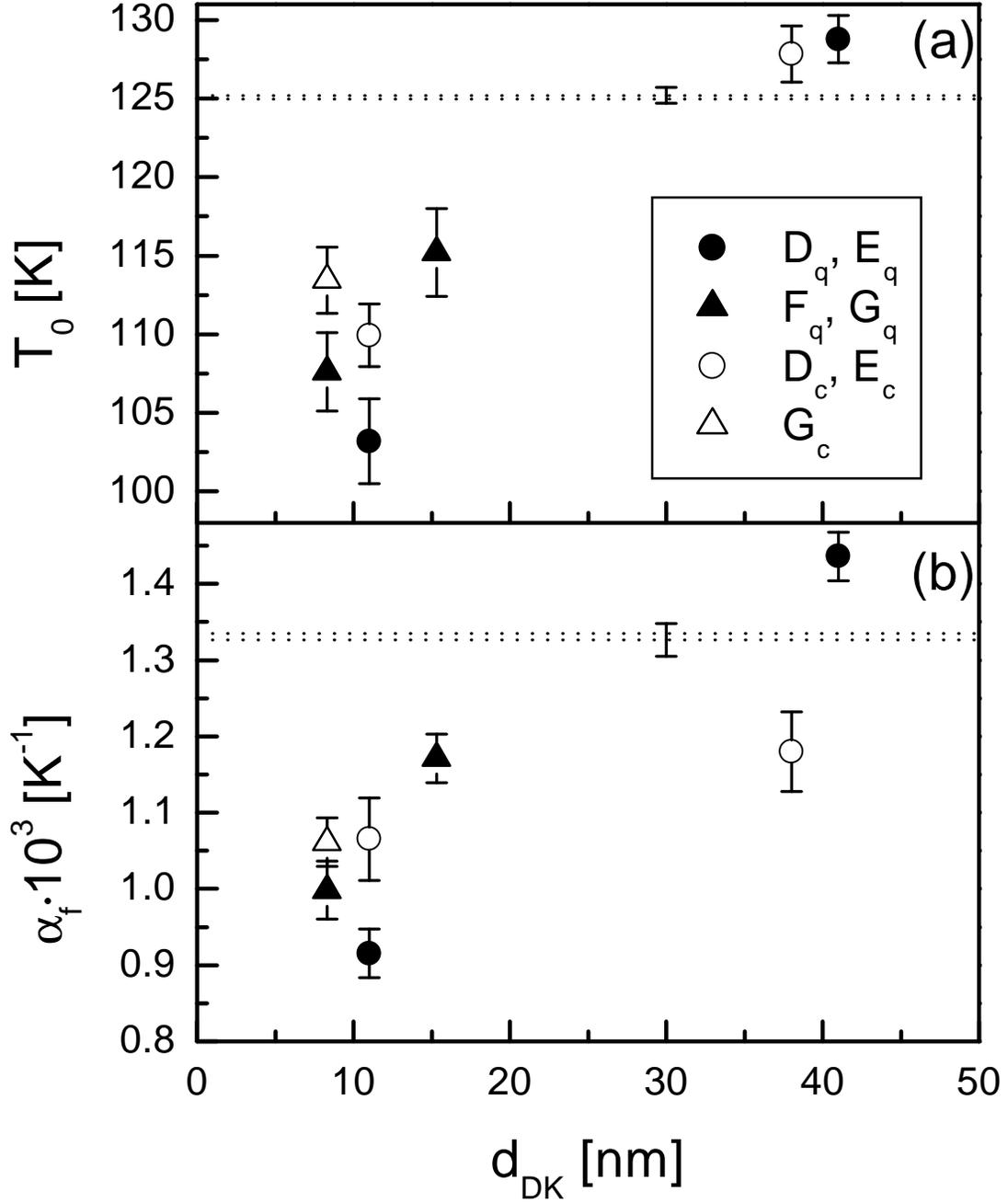}
\caption{(a) Vogel temperature $T_{\rm 0}$ and (b) thermal expansion coefficient of the free volume $\alpha_{\rm
f}=1/B$ of PDMS in the bulk and as grafted layers vs sample thickness $d$ as obtained from fits of the VFT-equation
(Eq. \ref{vfteq}) to the relaxation rates of the $\alpha$-relaxation. For grafted layers the results for two molecular
weights and for different thermal treatments of the samples are given: Full symbols denote measurements on samples
which were quenched prior to the dielectric measurements, open symbols refer to measurements on samples which were
cooled slowly. Circles represent the molecular weight $M_{\rm w}$=170000 g mol$^{\rm -1}$, triangles represent $M_{\rm
w}$=96000 g mol$^{\rm -1}$. Bulk values are given by dotted lines; for values of parameters and sample codes see Table
\ref{vfttab}.}\label{fig:4}
\end{center}
\end{figure}

\vspace{.5cm}\hspace{5cm}{\it Please insert Figure \ref{fig:4}}\vspace{.5cm}

\noindent Figure \ref{fig:4} displays the dependence of the Vogel temperature $T_{\rm 0}$ and of the thermal expansion
coefficient of free volume, $\alpha_{\rm f}=1/B$ as obtained from fits of Eq. \ref{vfteq} to the temperature dependence
of the relaxation time $\tau$. For the fits $\tau_{\rm \infty}$ was fixed to the value obtained from bulk measurements
($log(\tau_{\rm \infty})=14.5$). Thus, only $B$ and $T_{\rm 0}$ were adjusted to reproduce the observed values
\cite{FukaoPRE301}. Alternatively, fits were carried out by adjustment of all three parameters. The tendencies given in
Fig. \ref{fig:4} are independent of the chosen fitting procedure. It is found that both, the Vogel temperature $T_{\rm
0}$ and the thermal expansion coefficient $\alpha_{\rm f}$ of the free volume decrease when the film thickness is
reduced. Table \ref{vfttab} summarizes the values for both, $T_{\rm 0}$ and $\alpha_{\rm f}$ for various samples.
Additionally, values for the glass transition temperature $T_{\rm g}$ as the temperature where the relaxation rate is
equal to 100 s are given.

\vspace{.5cm}\hspace{5cm}{\it Please insert Table \ref{vfttab}}\vspace{.5cm}

\noindent In the framework of the free volume theory by Cohen and Turnbull \cite{CohenTurn59} the relaxation time
$\tau_{\rm \alpha}$ of the $\alpha$-relaxation is given by:
\begin{equation}\label{taualpha}
\tau_{\rm \alpha}=\tau_{\rm \infty}\,\exp\,(\frac{b}{f}),
\end{equation}
where $b$ is a constant and $f$ is the fractional free volume. For any film thickness $d$, the latter is related to the
Vogel temperature $T_{\rm 0}$ and the expansion coefficient $\alpha_{\rm f}$ via:
\begin{equation}\label{freevol}
f(d;T)=\alpha_{\rm f}(d)(T-T_{\rm 0}(d)).
\end{equation}
Recently, Fukao and co-workers \cite{FukaoPRE200} reported that the $\alpha$-relaxation in thin spincast films of
polystyrene is faster than in the bulk. In frame of the free volume theory, the decrease of both, $\alpha_{\rm f}$ and
$T_{\rm 0}$ in thin grafted films leads to two competing trends concerning the relaxation time $\tau_{\rm \alpha}$:
while the former tends to slow down the $\alpha$-relaxation, the latter can be regarded as an increase of the
relaxation rate of the $\alpha$-process. From these considerations it was concluded that for sufficiently high
temperatures the relaxation dynamics in thin films should become slower than in the bulk. In the present study this
crossover temperature could be estimated to be around 180 K. Since this value is above the temperature of
crystallization, for thin films of amorphous PDMS only a faster $\alpha$-relaxation than that in the bulk is to be
expected.

\noindent Qualitatively, the glass transition temperature $T_{\rm g}$ should behave similar to the Vogel temperature
$T_{\rm 0}$ (see Tab. \ref{vfttab}). It has been reported recently that $T_{\rm g}$ in thin grafted films of
polystyrene increases for films much thicker than the radius of gyration \cite{FryerJCP01}. This result resembles to
our finding for the film with $d>R_{\rm g}$, where the $\alpha$-relaxation is slowed down and $T_{\rm g}$ is
consequently increased with respect to the bulk. However, in \cite{FryerJCP01} no results for $T_{\rm g}$ are given for
the case $d<R_{\rm g}$ as it is investigated in the present study and where a decrease of $T_{\rm g}$ is found. The
opposed findings of an increase and decrease of $T_{\rm g}$, respectively may be due to differences of the investigated
polymers. While polystyrene has rather stiff chains, poly(dimethylsiloxane) is characterized by a high chain
flexibility.

\section{Conclusions}\label{conclus}
\noindent The dynamics of the $\alpha$- and $\alpha_{\rm c}$-relaxation of PDMS in the bulk and as grafted films have
been investigated by use of dielectric relaxation spectroscopy. The results of this study can be summarized as follows:
\begin{enumerate}
\item The evaluation of the temperature dependence of the $\alpha$-relaxation strongly suggests that its dynamics in
grafted films of PDMS depends on the ratio of the film thickness $d$ to the radius of gyration $R_{\rm g}$ ({\it chain
confinement}). In case of the $\alpha$-relaxation, this dependence of the molecular dynamics on the film thickness can
be well distinguished from the dependence on the thermal history of the samples. \item In films with $d<R_{\rm g}$ the
relaxation rate of the $\alpha$-relaxation increases by up to two orders of magnitude compared to the bulk. \item For
films with a thickness larger than $R_{\rm g}$ the relaxation rate of the $\alpha$-relaxation is slowed down by one
order of magnitude compared to that of the bulk. This slowing down is considered as a prevailing of interactions
between the polymer and the substrate over the influence of spatial (chain) confinement. \item While the dielectric
strength $\Delta\varepsilon$ of the $\alpha$-relaxation in films with $d>R_{\rm g}$ is comparable to that of the bulk,
it is reduced by up to 60\% in the thinner films. This decrease is explained by a partial immobilization of chain
segments. \item The Vogel temperature $T_{\rm 0}$ is decreased by up to 20 K in films with $d<R_{\rm g}$, whereas it is
slightly increased in films with $d>R_{\rm g}$. \item Within the framework of free volume theory, the thermal expansion
coefficient of free volume $\alpha_{\rm f}$ is found to decrease in films of a thickness below $R_{\rm g}$. \item For
the $\alpha_{\rm c}$-relaxation which is related to the dynamics of the amorphous fraction of PDMS in partial
crystalline samples, any possible effect of the confinement is outweighed by the dependence of this relaxation process
on the thermal history of the samples.
\end{enumerate}

\noindent Qualitatively, the finding concerning the relaxation time of the $\alpha$-relaxation in grafted films with
$d<R_{\rm g}$ (Fig. \ref{fig:2}) is in accord with results for small molecules confined to pores or channels with a
diameter of 0.5 nm to 10 nm: with decreasing thickness, $\tau_{\rm \alpha}$ is shifted to smaller values. Moreover, the
results 2. and 3. lead to the conjecture of the existence of a critical thickness $d_{\rm c}$ at which the dynamics of
the $\alpha$-relaxation changes from that in the bulk to that in thin films. In case of confined small molecules, this
result could be explained unambiguously by an interference of the size of confinement with a growing correlation length
$\xi$ (in the order of 1...10 nm) inherent to the glass transition \cite{ArndtPRL97}. However, in case of grafted PDMS
deviations from the $\alpha$-relaxation time in the bulk appear already at film thicknesses of 15 nm (Fig.
\ref{fig:2}), whereas the molecular dynamics in films of 40 nm thickness resembles well to that of the bulk except a
slight decrease of the relaxation rate. From dielectric and temperature modulated calorimetric (TMDSC) measurements on
PDMS ($M_{\rm w}$=1400 g mol$^{\rm -1}$) confined to nanoporous sol-gel glasses with pore diameters ranging from 2.5 nm
to 20 nm the length scale for {\it cooperative} fluctuations is estimated to be between 5 nm and 7.5 nm at the glass
transition temperature \cite{DKBuch602}. But also in pores of a diameter of 20 nm an increased relaxation rate of the
$\alpha$-relaxation of PDMS is observed with respect to that of the bulk. This is in agreement to our findings for
films of a comparable film thickness, i.e. $d<R_{\rm g}$, whereas the slowing down of the $\alpha$-relaxation in
thicker films is attributed to adsorption. From these results it has to be concluded that the onset of confinement
effects for the $\alpha$-relaxation appears at sizes of the confinement well above the length scale for cooperative
fluctuations.

\section*{Acknowledgement}\label{acknow}
\noindent We gratefully acknowledge support of the DFG within Sonderforschungsbereich 294, TP G6.

%
%
%
\newpage
\begin{table}
\caption{Values of various parameters for bulk samples (A-C) and grafted layers (D-G) of PDMS of two different
molecular weights. $M_{\rm w}$ denotes the mass average of the molecular weight, $\phi$ is the volume fraction of
grafted short chains to decouple the grafted long chains from the substrate. $R_{\rm g}$ is the radius of gyration and
$d_{\rm diel}$ and $d_{\rm ellip}$ are the values of the film thickness as determined from measurements of the
capacitance and from ellipsometric measurements, respectively. $\sigma$ denotes the grafting density and $b$ the
distance between grafting sites.} \label{geom} \vspace{3ex}
\begin{tabular}{cccccccc}
\colrule \colrule sample code & $M_{\rm w}$ [g mol$^{\rm -1}$] & $\phi_{\rm (M_w=5000\: g\: mol^{-1})}$& $R_{\rm
g}$\footnote{$R_{\rm g}$ was determined from extra- and interpolation of values given in \cite{HigginsPolym79}.} [nm] &
$d_{\rm
diel}$ [nm] & $d_{\rm ellip}$ [nm] & $\sigma$ & $b$ [nm]\\
\colrule
A&1.7$\cdot$10$^{\rm 5}$&---&35.6&3.2$\cdot$10$^{\rm 4}$&---&---&---\\
B&9.6$\cdot$10$^{\rm 4}$&---&20.5&5$\cdot$10$^{\rm 4}$&---&---&---\\
C&3.6$\cdot$10$^{\rm 3}$&---&1.6&1.4$\cdot$10$^{\rm 4}$&---&---&---\\
\colrule
D&1.7$\cdot$10$^{\rm 5}$&0.08&35.6&40$\pm$2&---&0.14&2.6\\
E&1.7$\cdot$10$^{\rm 5}$&0.04&35.6&11&8&0.038&5.1\\
\colrule
F&9.6$\cdot$10$^{\rm 4}$&0&20.5&15&---&0.095&3.3\\
G&9.6$\cdot$10$^{\rm 4}$&0.10&20.5&8&6.9&0.051&4.4\\
\colrule \colrule
\end{tabular}
\end{table}
\mbox{} \thispagestyle{empty}
\newpage
\mbox{} \thispagestyle{empty}
\begin{table}
\vspace{2cm} \caption{Values for parameters of VFT-fits (Eq. \ref{vfteq}) obtained by the fitting procedure described
in the text. $T_{\rm 0}$ denotes the Vogel temperature. The glass transition temperature $T_{\rm g}$ is determined as
that temperature where the VFT-fit yields a relaxation time $\tau$ of 100 s. The fit parameter $A$ in the VFT-fits was
fixed at 14.5. $\alpha_{\rm f}=B^{\rm -1}$ denotes the thermal expansion coefficient of the free volume.}
\label{vfttab}\vspace{3ex}
\begin{tabular}{ccccc}
\colrule \colrule sample code\footnote{The indices q and c refer to quenched and cooled
samples, respectively.}&$T_{\rm 0}$ [K]&$\alpha_{\rm f}\cdot$10$^{\rm 3}$ [K$^{\rm -1}$]&$T_{\rm g}$ [K]\\
\colrule
A$_{\rm q}$&125&1.33&145\\
B$_{\rm q}$&125&1.33&145\\
C$_{\rm q}$&123&1.32&143\\
\colrule
D$_{\rm q}$&127&1.34&147\\
D$_{\rm c}$&124&1.08&149\\
E$_{\rm q}$&102&0.91&131\\
E$_{\rm c}$&104&0.90&133\\
\colrule
F$_{\rm q}$&117&1.20&139\\
G$_{\rm q}$&111&1.11&135\\
G$_{\rm c}$&113&1.03&138\\
\colrule \colrule
\end{tabular}
\end{table}
\setlength{\topmargin}{-3cm} \setlength{\textheight}{30cm}
\end{document}